\documentclass[%
 reprint,
 amsmath,amssymb,
]{revtex4-2}

\usepackage{graphicx}
\usepackage{physics}
\usepackage{xcolor}

\DeclareMathOperator*{\argmin}{arg\,min}
\usepackage{dcolumn}
\usepackage{bm}

\usepackage{hyperref}

\begin{document}

\setlength{\parindent}{0pt}

\preprint{APS/123-QED}

\title{A New \textsc{Herwig}7 Underlying Event Tune: from RHIC to LHC Energies}
\author{Umar Sohail Qureshi}
 \email{uqureshi@cern.ch}%
\author{Raghav Kunnawalkam Elayavalli}

\affiliation{Department of Physics and Astronomy, Vanderbilt University, Nashville, Tennessee, USA}%

\author{Luke Mozarsky}
\email{luke.mozarsky@cern.ch}
\author{Helen Caines}
\author{Isaac Mooney}

\affiliation{Department of Physics, Yale University, New Haven, Connecticut, USA}

\date{\today}

\begin{abstract}
 We present parameter sets corresponding to new underlying event tunes for the {\textsc{Herwig}7.3} Monte Carlo event generator. The existing \textsc{Herwig} tunes are in good agreement with LHC data, however, they are not typically designed for center-of-mass energies below $\sqrt{s}=300$ GeV. The tunes presented in this study can describe mid-rapidity data collected at the nominal RHIC energy of $\sqrt{s }=200$ GeV as well as higher center-of-mass energies utilized by experiments elsewhere, such as the LHC. The base ``New Haven" tune is developed by fitting minimum-bias simulations of proton-proton collisions to mid-rapidity identified hadron and jet data from the STAR experiment. The ``Nashville" tune includes a separate set of parameters developed by tuning to Tevatron proton-antiproton data at $\sqrt{s}=300$, $900$ and $1960$ GeV from CDF, and LHC proton-proton measurements from CMS at $\sqrt{s}=7$ TeV, in addition to the STAR measurements. Both new tunes demonstrate significant improvements over the recommended default tune currently included in the latest version of \textsc{Herwig} for minimum bias production. As such, we advocate using these tunes for future simulation studies at mid-rapidity by the experimental collaborations at RHIC (STAR and sPHENIX) and the LHC (ATLAS, ALICE, CMS).
\end{abstract}

\maketitle











\section{Introduction}
\label{introduction}

Monte Carlo (MC) event generators, which simulate collisions involving relativistic lepton-lepton, lepton-hadron, and hadron-hadron interactions play a critical role in high-energy particle and nuclear physics. Theoretically, MC models help test our fundamental understanding of the Standard Model, particularly quantum chromodynamics (QCD), providing a framework for the initial and final states of collision systems. Experimentally, MCs are a vital part of the simulation process, aiming to reproduce realistic spectra used for determining detector acceptance and resolution corrections and studying systematic effects in experimental data. Popular event generators, such as \textsc{Pythia} \cite{Bierlich:2022pfr, Sjostrand:2006za} and \textsc{Herwig}/\textsc{Herwig}++ \cite{Bewick:2023tfi, Bahr:2008pv}, are extensively used to simulate collisions at the Large Hadron Collider (LHC) and the Relativistic Heavy-Ion Collider (RHIC). These simulations include various QCD physics processes, which separate a single collision into a perturbative hard scattering regime and evolution via a parton shower, incorporating both Initial State Radiation (ISR) and Final State Radiation (FSR). Additionally, they account for non-perturbative components such as hadronization, the underlying event (UE), and multi-parton interactions (MPI).

The aforementioned MC event generators use multiple free parameters that are determined by fits to experimental data. The \textsc{Herwig}7 MC event generator discussed in this study has undergone numerous tuning exercises over the years and is successful in describing data at the LHC \cite{CMS:2020dqt, Bellm:2019zci}. However, significant discrepancies remain when describing data from collisions at lower center-of-mass energies. These disagreements are mainly attributed to incorrect modeling of the low-energy QCD interactions in the underlying event, which arise from the extrapolation of the center-of-mass energy used. For \textsc{Herwig}7, several tuning exercises were conducted using LHC data. Currently, the set of parameters from Ref. \cite{Bellm:2019icn} is the recommended option for the simulation of minimum bias and underlying event data. This set of parameters is obtained in the following manner. First, the authors tune MPI parameters explicitly responsible for determining the soft/hard emission threshold, which has a power-law dependence on $\sqrt{s}$. These parameters are tuned to data taken with center-of-mass collision energy ranging from $\sqrt{s} = 200$ GeV to $\sqrt{s} = 13$ TeV, one energy at a time. The authors then fit the aforementioned power law to the emission thresholds produced by these tunes to extract MPI parameters that, in principle, describe multiple-parton interactions well across the entire range of $\sqrt{s}$. This power law is forced to interpolate the points at $\sqrt{s} = 7, 13$ TeV, which the authors note leads to improved tune performance at higher center-of-mass energies while resulting in a sub-optimal description of data at $\sqrt{s} = 200$ GeV. Finally, the remaining MPI model and color reconnection parameters are tuned to minimum bias and underlying event data taken at $\sqrt{s} = 7, 13$ TeV. We find that this tune, in addition to other existing \textsc{Herwig} tunes, does not provide a satisfactory description of data with energy around $\sqrt{s}=200$ GeV. In particular, this limits the utility of \textsc{Herwig} as a tool for the STAR and sPHENIX experiments at RHIC.

In this work, we use underlying event multiplicity data from RHIC at the nominal center-of-mass energy $\sqrt{s}=200$ GeV, CDF at energies of 300, 900 and 1960 GeV, as well as LHC data at energies of 7 TeV, to produce two tunes with improved performance at $\sqrt{s} = 200$ GeV while maintaining a good description of collisions at higher center-of-mass energies. A similar dedicated \textsc{pythia}8 tune was developed in Ref. \cite{pythia8}, and was seen to significantly improve the description of experimental data.

This paper is organized as follows: Section \ref{sec:tuningmethod} details the Professor toolkit and its parameterization-based tuning procedure. Section \ref{sec:samplesandsim} describes the data and its implementation through the RIVET framework. Section \ref{sec:ourtunes} presents the results for our new tunes. Section \ref{sec:comps} compares the new \textsc{Herwig}7 tune predictions with selected data distributions. We conclude with a discussion in Section \ref{sec:discussion} that summarizes our findings.

\section{Tuning Methodology}
\label{sec:tuningmethod}
For this study, we adopt a parameterization-based tuning procedure via the Professor (v2.4.2) toolkit \cite{Buckley:2009bj}. This process involves sampling our parameters of interest multiple times within a specified range, and, for each sample of parameters, generating \textsc{Herwig}7 Monte Carlo (MC) events. The resulting predictions of these MC simulations are then compared to data using RIVET \cite{Bierlich:2019rhm}. For each bin of each data observable, Professor constructs a third-degree polynomial to parameterize the \textsc{Herwig}7 predictions as a function of the $P$ tuning parameters, with numerically computed coefficients. Finally, a $\chi^2$ fit of the polynomial to the data is performed using the Minuit framework ~\cite{James:1975dr} which identifies the optimal set of parameters.

The \textsc{Herwig} Monte Carlo (MC) event generation framework consists of both perturbative and non-perturbative components. The hard component of an event refers to the particles produced through the showering process of the partons that emerge from the primary hard 2-$n$ scattering. In contrast, the underlying event comprises particles originating from the hadronization of beam remnants and MPI, as well as their related initial-state and final-state radiation. Beam-beam remnants are made up of hadrons that arise from the fragmentation of remnant partons, which participate minimally in the momentum transfer during the collision. MPI refers to additional parton-parton scatterings (2-to-2 interactions) occurring within the same hadronic collision but characterized by a lower transverse momentum compared to the primary hard scattering event. 

In addition to hard-scattering processes, several other mechanisms contribute to the inelastic cross-section in hadron-hadron collisions. These include single-diffraction dissociation (SD), double-diffraction dissociation (DD), and central diffraction (CD). For a detailed description of these mechanisms, see Ref. \cite{gieseke2016diffractionherwig}.

The term minimum bias (MB) refers to a broad class of events that are selected with minimal requirements on the detector activity, effectively capturing a large fraction of the overall inelastic cross section. MB data are frequently used in studies of the underlying event (UE) because they provide an inclusive sample of inelastic collision events.

As a starting point, we use the default tune \cite{Bellm:2019icn} (which we label as ``default") and an updated set of NNPDF3.0 \cite{NNPDF:2014otw} leading-order parton distribution functions (PDFs). The tuned parameters are described as follows. The \textsc{Herwig}7 MPI model is determined primarily by the two parameters $p_{\perp}^{\mathrm{min}}$ and $\mu^2$. Emissions with $p_\perp> p_{\perp}^{\mathrm{min}}$ are considered as hard, while those below are considered soft. The first set of our tuning parameters $(p_{\perp,0}^{\mathrm{min}}, b,c)$ governs the energy-scaling of $p_{\perp}^{\mathrm{min}}$, which follows a power-law function
\begin{align}\label{ptmin_extrap}
p_{\perp}^{\mathrm{min}}\left(\sqrt{s}\right) &= p_{\perp,0}^{\mathrm{min}}\left(\frac{b+\sqrt{s}}{E_0}\right)^c
\end{align}
Where $E_0$ is the reference energy scale. The parameter $E_0$ is redundant, and thus we set $E_0 = 200$ GeV rather than its default value of $7$ TeV. Meanwhile, the parameter $\mu^2$ is the (squared) inverse of the proton radius, i.e. the transverse spatial extension of the partonic point cloud that enters the collision. Another set of parameters ($N_\mathrm{ladder}, B_\mathrm{ladder}$) governs the production of soft particles in a ladder graph. The expected number of particles $\expval{N}$ is parametrized such that the number $N$ of the particles in the the soft ladder is drawn from a Poisson distribution with mean $\expval{N}$ given as
\begin{align}
    \expval{N} &= N_\mathrm{ladder}\ln \left[\frac{(p_{r1}+p_{r2})^2}{m_\mathrm{rem}^2}\right] + B_\mathrm{ladder}
\end{align}
Where $p_{r1,2}$ are the momenta of the incoming beam remnants and $m_\mathrm{rem}$ is the constituent mass of the remnant. Following the guidelines in Ref. \cite{Bellm:2019icn}, we set the parameter $B_\mathrm{ladder}$ to zero and only tune the $N_\mathrm{ladder}$ parameter. 

Meanwhile, the diffractive ratio parameter $R_\mathrm{diff}$ gives the fraction of diffractive cross-section with respect to the nondiffractive cross-section. The last of our tuning parameters is the color reconnection probability $p_\mathrm{reco}$, which determines the likelihood of rearranging the color topology between clusters or QCD dipoles to minimize the total invariant cluster mass. A more detailed description of the above tuning parameters is given in Ref. \cite{Bellm:2019icn}. All the tuning parameters and their respective ranges are listed in Table \ref{tab:tuningparams}.
\begin{table}
        \begin{tabular}{c  c  c  c c c}
        \hline
        \textbf{\quad Setting\quad} &  \textbf{\quad Default\quad} & \textbf{\quad Our Tunes\quad} 
        \\
        \hline
        $E_0$ & 7 TeV & 200 GeV\\
        PDFSet & CT14LO & NNPDF30\\
        \hline
        \textbf{\quad Parameter \quad} &  \textbf{\quad Default \quad} & \textbf{ \quad Range \quad} 
        \\
        \hline
        $c$ &  0.31 & 0.15--0.50  \\

        $b$ & 622& 500--950  \\
  
        $p_{\perp,0}^{\mathrm{min}}$ &  2.82 & 2.00--5.00\\
    
        $\mu^2$ & 1.1 & 0.68--2.50 \\

        $N_\mathrm{ladder}$ & 0.6838 & 0.2--0.8 \\

        $p_\mathrm{reco}$ & 0.970 & 0.2--1.0 \\
        $R_\mathrm{diff}$ & 0.187 & 0.18--0.40 \\
        \hline
        \end{tabular}
        \caption{Comparison of \textsc{Herwig}7 settings and tuning ranges for parameters between the default tune and the tunes presented in this study.}
        \centering
        \label{tab:tuningparams}
\end{table}
For the tuning procedure, we note that for 7 tuning parameters, a third-degree polynomial has 120 coefficients, thereby requiring a minimum of 120 samples in parameter space. To oversample by at least a factor of two as recommended by the Professor authors \cite{Buckley:2009bj}, we require 240 points. Hence, we sample 300 values of the tuning parameters in Table \ref{tab:tuningparams} within the specified ranges to be used as anchor points in the generator response polynomial parametrization. For each sampling, we generate 10 million minimum bias inelastic events to ensure the MC statistics are sufficient as compared to the data. The optimal vector of tuned parameters $\boldsymbol{\theta_0}$ is determined by minimizing the weighted $\chi^2$, where the optimization is over all parameter vectors $\boldsymbol{\theta}\in \Theta$:
\begin{align}
    \boldsymbol{\theta_0}
    &= \argmin_{\boldsymbol{\theta} \in \Theta} \chi^2(\boldsymbol{\theta})
\end{align}
Where the $\chi^2$ for a single parameter vector $\boldsymbol{\theta}$ is computed as:
\begin{align}
   \chi^2(\boldsymbol{\theta}) &= \sum_{i\in I} w_i (f_i(\boldsymbol{\theta}) - d_i) C_i^{-1} (f_i(\boldsymbol{\theta}) - d_i)
\end{align}
The summation index $I$ is over all data points $d_i$; $w_i$ is the weight for each data point; $f_i(\boldsymbol{\theta})$ is the corresponding interpolated {\textsc{Herwig}} response parameterized by a vector $\boldsymbol{\theta}$. $C_i$ is the covariance matrix considering just experimental uncertainties, it is assumed there are no significant bin-by-bin correlations. This assumption is reasonable because, for most experimental measurements used in tuning, the systematic uncertainties are typically either fully uncorrelated or treated in a way that averages out correlations across bins.

\section{Data and Simulation}
\label{sec:samplesandsim}

For this tuning exercise, we utilize mid-rapidity data in proton-proton collisions from the STAR and CMS experiments, as well as $p\bar{p}$ collisions from the CDF experiment. We choose datasets for the tuning procedure  to represent three main categories of measurements: identified particle spectra, which track specific particle types and their invariant yields; event multiplicities, which count the number of particles produced in collisions; and jet substructure measurements, which examine the radiation patterns in phase space of the hard scattered partons. These categories and the specifics of experiments used are summarized in Table \ref{tab:data}. 

The identified hadron spectra measurements from STAR \cite{STAR:2006xud} focus on $\pi^\pm$, $p$, and $\overline{p}$ yields at RHIC. These spectra are measured near midrapidity with $|y| < 0.5$ over the range $0.3 < p_\perp < 10$ GeV, utilizing particle identification techniques based on ionization energy loss and its relativistic rise in the Time Projection Chamber, along with time-of-flight measurements in the STAR experiment. The spectra are particularly sensitive to the UE since the UE contributes substantially to the production of low- and intermediate-$p_\perp$ particles.


\begin{figure}
    \centering
    \includegraphics[width=0.85\linewidth]{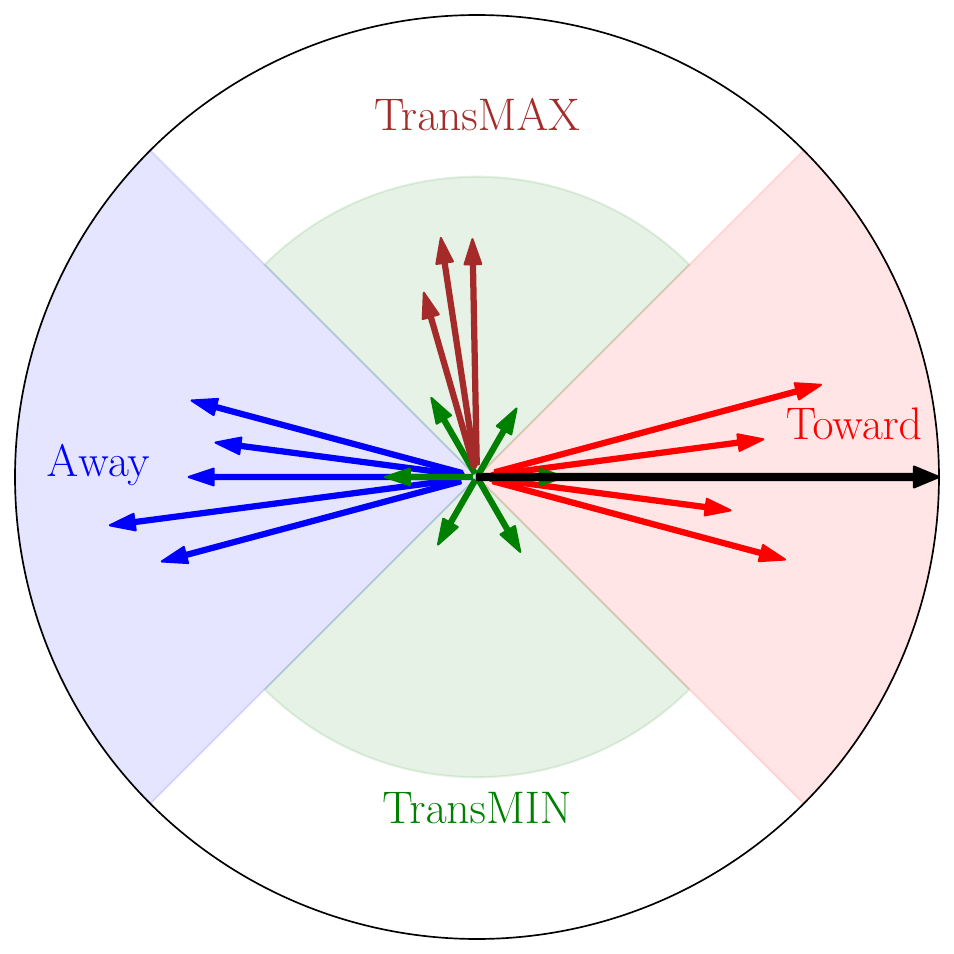}
    \caption{An illustration of a hadron-hadron collision in which a hard parton–parton collision has occurred, and the leading object is taken to be the charged particle of largest $p_\perp$ in the event. The black arrow in the toward direction indicates the leading object direction.}
    \label{fig:jettopo}
\end{figure}

For the UE multiplicity data from STAR \cite{STAR:2019cie}, CDF \cite{CDF:2015txs}, and CMS \cite{CMS:2011qzf}, the charged particle with the highest transverse momentum is defined as the leading object. Charged particles with $p_\perp > 0.5$ GeV and $\abs{\eta}<0.8$ are then used to describe the UE for the CDF and CMS measurements while $ p_\perp > 0.2$ GeV and $\abs{\eta}<0.4$ are used for the STAR data. The leading object direction is used to divide the detector $\eta{-}\phi$ space into four distinct regions on an event-by-event basis. The ``Toward" and ``Away" regions are defined relative to the leading object direction. The two transverse regions are then labeled as TransMAX or TransMIN, as depicted in Figure \ref{fig:jettopo}. The TransMAX (and TransMIN) regions are defined on an event-by-event basis as the regions with the greatest (and least) number of charged particles, or larger (or smaller) $p_\perp$ sums. The average transverse density TransAVE is the average of the TransMAX and the TransMIN densities while TransDIF is the the difference between TransMAX and TransMIN densities. Events with hard initial or final-state radiation, often result in a third jet in the TransMAX region. Both the TransMAX and TransMIN regions receive contributions from multi-parton interactions and beam-beam remnant components. The TransMIN region is very sensitive to the multi-parton interactions and beam-beam remnant components of the UE, while the TransDIF is very sensitive to initial and final-state radiation.

STAR measurements \cite{STAR:2020ejj} of the shared momentum fraction $z_g$ and the groomed jet radius $R_g$, as defined by the SoftDrop algorithm, are also used. These substructure observables are measured differentially for jets with varying resolution parameters from $R = 0.2$ to $0.6$ within a transverse momentum range of $15 < p_{\perp} < 60$ GeV. Finally, inclusive measurements of the invariant and SoftDrop groomed jet mass \cite{STAR:2021lvw} by the STAR experiment are considered. All jet substructure measurements are fully corrected for detector effects and reported differentially in both the jet transverse momentum and jet radius parameters. Particle within the momenta, energy, and rapidity ranges $0.2 < p_\perp < 30$ GeV, $0.2 < E_\perp < 30$ GeV, and $|\eta| < 1$ are included in the analysis. While jet substructure observables are not inherently very sensitive to UE, the authors of Ref. \cite{pythia8} saw incidental improvements from including them. This is likely because the inclusion of jet substructure helps fine-tune the extrapolation of $p_{\perp}^{\mathrm{min}}$ since the energy cutoff dictates the particle multiplicity in a jet. As a whole, the selected data effectively cover the available phase space and offer a robust metric to evaluate the tune's performance in capturing both the non-perturbative and perturbative components of hadronic collisions. 

The MC runs for all energies and beam types over all samples are then analyzed using the RIVET framework. The resulting output yoda files are processed by the Professor code to determine the minimum $\chi^2$. RIVET analyses for each of the measurements mentioned above are publicly available at \url{www.github.com/lmozarsky/herwig7-rhic-tune}.


\begin{table*}
\centering
\begin{tabular}{c c c c c c c}
\hline
\textbf{Experiment} & $\mathbf{\sqrt{s} \textbf{ (GeV)}}$ & \textbf{Observable} & \textbf{Nashville} & \textbf{New Haven} &\textbf{Figure} & \textbf{Reference}  \\ \hline
STAR & 200 & $\pi^\pm$ and $p\overline{p}$ cross sections vs $p_T$ & \checkmark & \checkmark & \ref{fig:STAR_comps}, \ref{fig:STAR_2006_PID} & \cite{STAR:2006xud}   \\  
STAR & 200 & Charged particle multiplicities and $p_T$ & \checkmark & \checkmark & \ref{fig:STAR_comps}, \ref{fig:STAR_2019_UE}  & \cite{STAR:2019cie}\\
CDF & 300, 900, 1960 & Charged particle density and $\sum p_T$ & \checkmark & -- & \ref{fig:CDF_1960}, \ref{fig:CDF_900}, \ref{fig:CDF_300} &  \cite{CDF:2015txs}   \\
CMS & 7000 & Charged particle density and $\sum p_T$ & \checkmark & --  & \ref{fig:CMS_7000}& \cite{CMS:2011qzf}\\
STAR & 200 & SoftDrop groomed jet substructure $(z_g, R_g)$ & \checkmark & \checkmark & \ref{fig:STAR_comps}, \ref{fig:softDropgroomedjet} & \cite{STAR:2020ejj}    \\  
STAR & 200 & Invariant and groomed jet mass  & \checkmark & \checkmark & \ref{fig:STAR_comps}, \ref{fig:softDropgroomedjet} &\cite{STAR:2021lvw}  \\ 
\hline
\end{tabular}
\caption{Summary of experiments and observables used in the new tunes.}
\label{tab:data}
\end{table*}

\section{New \textsc{Herwig}7 Tunes}\label{sec:ourtunes}

\begin{table}
        \begin{tabular}{c c c c}
            \hline
                \textbf{ Parameter } &  \textbf{ Default } & \textbf{  Nashville }  &  \textbf{New Haven }\\
        \hline
        $c$ &  0.31 & 0.36 & 0.23  \\

        $b$ & 622& 820 & 551 \\
  
        $p_{\perp,0}^{\mathrm{min}}$ &  2.82 & 2.99 & 3.01\\
    
        $\mu^2$ & 1.10 & 1.53 & 1.11 \\

        $N_\mathrm{ladder}$ & 0.6838 & 0.3966 & 0.5690 \\

        $p_\mathrm{reco}$ & 0.970 & 0.720 & 0.522 \\
        $R_\mathrm{diff}$ & 0.187 & 0.299 & 0.262 \\
        \hline
        \end{tabular}
        \caption{Comparison of \textsc{Herwig}7 tuned parameters between the default and the new Nashville and New Haven tunes.}
        \centering
        \label{tab:tunedparams}
\end{table}

\begin{figure*}
\centering
\begin{minipage}{.33\textwidth}
  \centering
  \includegraphics[width=\linewidth]{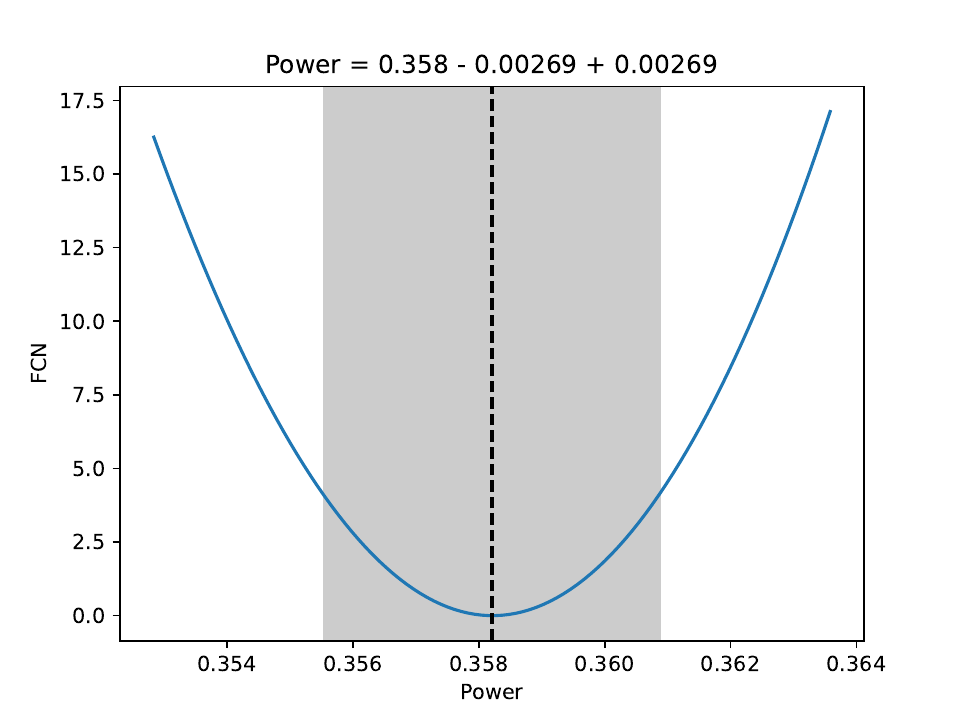}
\end{minipage}%
\begin{minipage}{.33\textwidth}
  \centering
  \includegraphics[width=\linewidth]{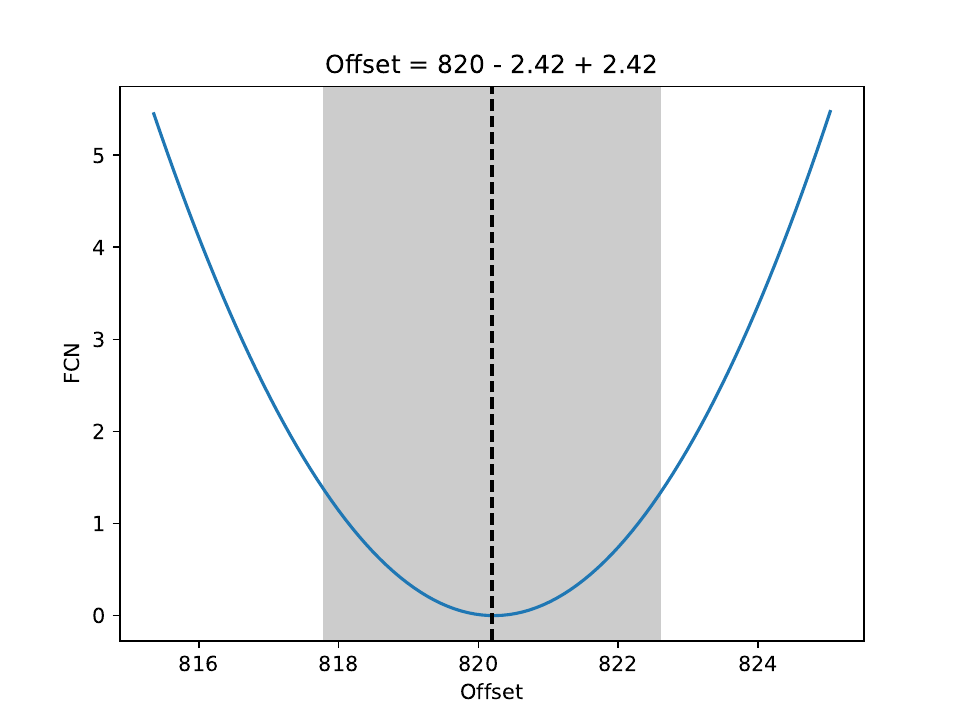}
\end{minipage}
\begin{minipage}{.33\textwidth}
  \centering
  \includegraphics[width=\linewidth]{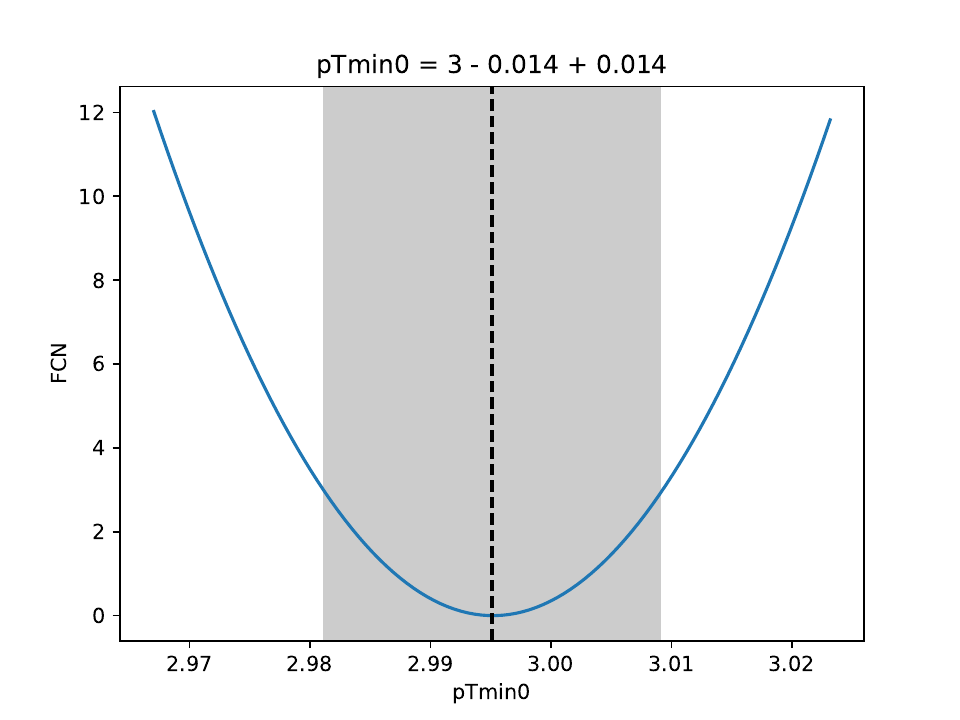}
\end{minipage}

\begin{minipage}{.33\textwidth}
  \centering
  \includegraphics[width=\linewidth]{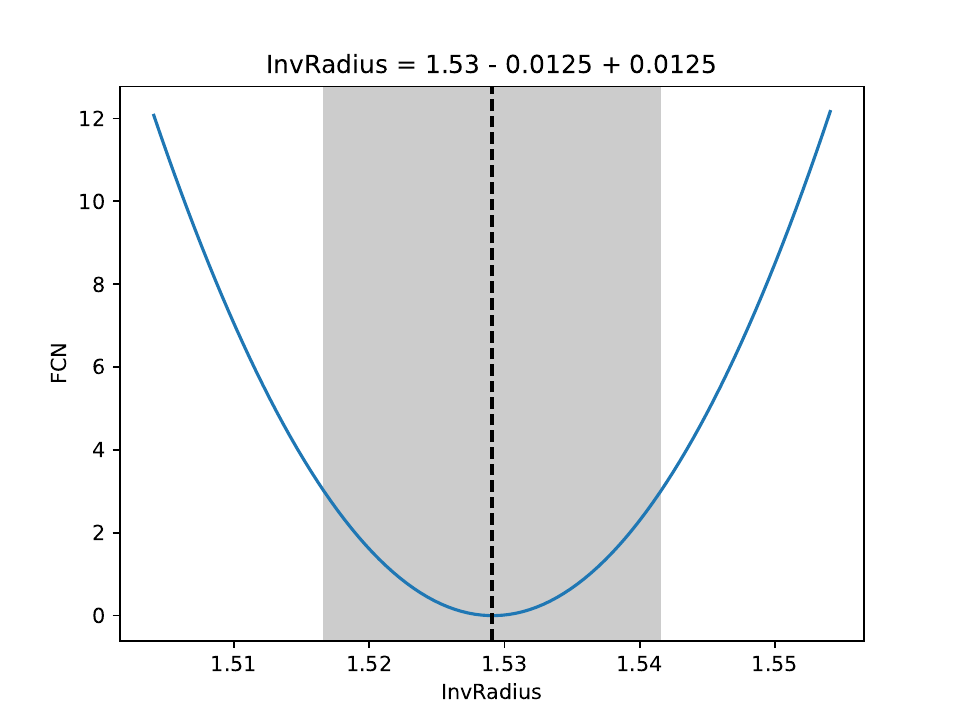}
\end{minipage}%
\begin{minipage}{.33\textwidth}
  \centering
  \includegraphics[width=\linewidth]{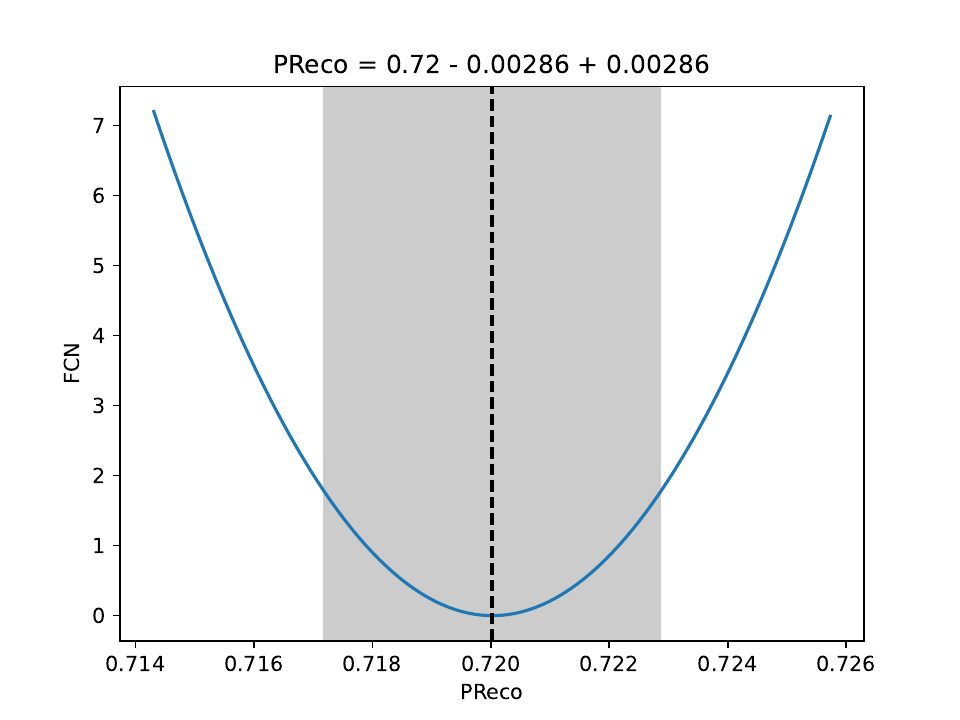}
\end{minipage}
\begin{minipage}{.33\textwidth}
  \centering
  \includegraphics[width=\linewidth]{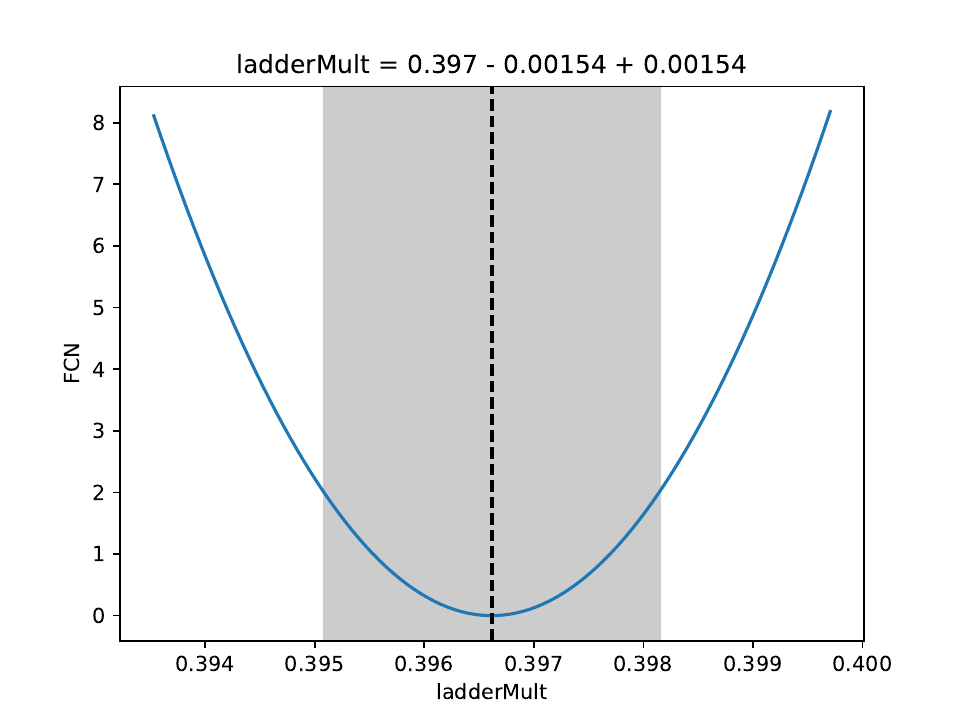}
\end{minipage}
\caption{Minimum-subtracted $\chi^2$ profiles of the Nashville tune parameters in the vicinity of the best-fit value. The grey-shaded regions indicate the $1$-$\sigma$ uncertainties in the optimal parameters. The parameters in the figure correspond to those in Table \ref{tab:tunedparams}, where \texttt{Power}, \texttt{Offset}, \texttt{pTmin0}, \texttt{InvRadius}, \texttt{PReco}, and \texttt{ladderMult} correspond to $c$, $b$, $p_{\perp, 0}^{\text{min}}$, $\mu^2$, $p_{\text{reco}}$, and $N_{\text{ladder}}$ respectively.}
\label{fig:chisquareprofiles}
\end{figure*}

Figure \ref{fig:chisquareprofiles} shows the $\chi^2$ profiles for each tuning parameter around the global minimum (offset such that $\chi^2 = 0$ at the minimum) for the Nashville tune. The $\chi^2$ statistic for the New Haven tune shows similar sensitivity to variations in the tuning parameters.

Table \ref{tab:tunedparams} lists the optimal values of the tuning parameters considered in the new Nashville and New Haven tunes. Generally, the errors on these central values derived from Minuit are negligible $(<1\%)$, do not affect the shape of the resulting MC responses, and are, therefore, omitted. However, they may be seen in the $\chi^2$ profile plots depicted in Figure \ref{fig:chisquareprofiles}. The best fit $\chi^2$ per degree of freedom $\chi^2 /\nu $ is $6971/1055 \approx 6.61$ for the Nashville tune and $1451/518 \approx 2.80$ for the New Haven tune.

\begin{figure}
    \centering
    \includegraphics[width=\linewidth]{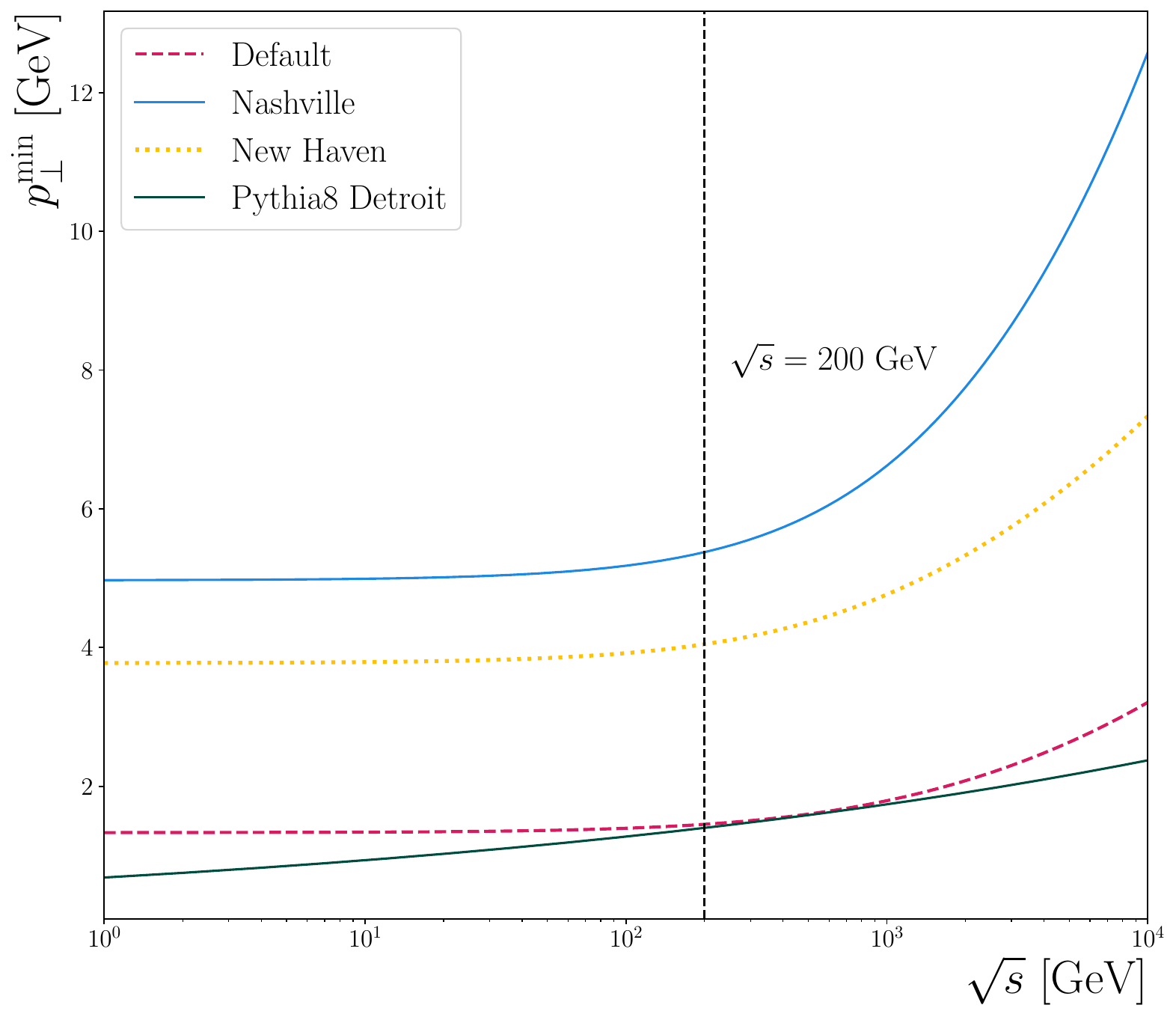}
    \caption{The energy extrapolations of $p_{\perp}^{\mathrm{min}}$ for the default, Nashville, New Haven, and \textsc{Pythia} Detroit tunes are shown as the solid blue, red, green, and purple lines, respectively. The black dashed line is drawn for reference at $\sqrt{s}=200$ GeV.}
    \label{fig:ptmin}
\end{figure}

Figure \ref{fig:ptmin} shows the extrapolation of $p_{\perp}^\mathrm{min}$ as dictated by Eq. \ref{ptmin_extrap} for the default, New Haven, Nashville, and \textsc{Pythia}8 Detroit tunes. We remind the reader that all tunes set the reference energy $E_0 = 200$ GeV except the default tune which uses $E_0 = 7$ TeV.
We observe that for both the Nashville and New Haven tunes, the new $p_{\perp}^{\mathrm{min}}$ value is roughly a factor of 2.5 to 3 larger compared to the default and the \textsc{pythia} 8 Detroit tunes up to $\sim 1$ TeV, after which the two new tunes sepearate slightly from each other towards higher $p_{\perp}^{\mathrm{min}}$. The Detroit and default tune extrapolations are very comparable at $\sqrt{s}=200$ GeV energies and diverge as we go up or down in collision energy. Between the Nashville and New Haven tunes, there is about a constant $20-25\%$ percent difference in the value of $p_{\perp}^{\mathrm{min}}$ from 0 to $200$ GeV, which then increases slightly at $\mathcal{O}(\mathrm{TeV})$. In general, the Nashville and New Haven tunes adopt a stricter approach to regulating hard QCD interactions compared to the default and the Detroit tunes. As the energy increases, the $p_{\perp}^{\mathrm{min}}$ variable in the Nashville and New Haven tunes grows much more rapidly, enforcing stricter cutoffs and leading to fewer hard particles, especially at high energies. We attribute much of the offset between the two new tunes and the Detroit and default tunes to the fact that the reference energy $E_0$ was changed from $7$ TeV to $200$ GeV. 

The comparison of \textsc{Herwig}7 tuned parameters between the default, Nashville, and New Haven tunes reveals several differences. 
For the parameter $c$, Nashville shows an increase (0.36) compared to the default (0.31), a rise of approximately 16.1\%. However, New Haven exhibits a significant decrease, with $c = 0.225$, which is about 27.4\% lower than the default and 37.5\% lower than Nashville. The parameter $b$ shows a large increase of about 31.8\% in Nashville (820) compared to the default (622). In New Haven, $b$ decreases to 551, which is roughly 33\% lower than in Nashville and 11.4\% lower than the default tune. Note that these changes manifest as the observed difference in the $p_{\perp}^{\text{min}}$ extrapolation.

The parameter $p_{\perp,0}^{\mathrm{min}}$ shows slight increases across both tunes, with Nashville (2.99) being about 6\% higher than the default (2.82) and New Haven (3.005) showing an approximately 6.6\% increase over the default. Again, this is evidence of the fact that our tunes predict a greater proportion of soft particle emissions than the default and Detroit tunes. The parameter $\mu^2$ increases significantly in Nashville (1.53) compared to the default (1.1), a rise of about 40\%, whereas New Haven shows a more modest increase to 1.11, only about 1\% higher than the default but 27.5\% lower than Nashville.

$N_{\mathrm{ladder}}$ shows a significant reduction in Nashville (0.3966) compared to the default (0.6838), representing a decrease of approximately 42\%. In New Haven, $N_{\mathrm{ladder}}$ rises to 0.569, which is still about 16.8\% lower than the default but 43.4\% higher than Nashville. For the parameter $p_{\mathrm{reco}}$, Nashville decreases to 0.720 from the default 0.970, a reduction of about 25.8\%. In New Haven, this value decreases further to 0.522, which is about 46.2\% lower than the default and 27.5\% lower than Nashville.

Finally, the parameter $R_{\mathrm{diff}}$ increases in both Nashville (0.299) and New Haven (0.262) compared to the default (0.187), representing increases of about 59.9\% and 40.1\%, respectively.

\section{Comparisons with Data}
\label{sec:comps}

\begin{figure*}
    \centering
    \begin{minipage}{0.495\textwidth}
        \includegraphics[width=\linewidth]{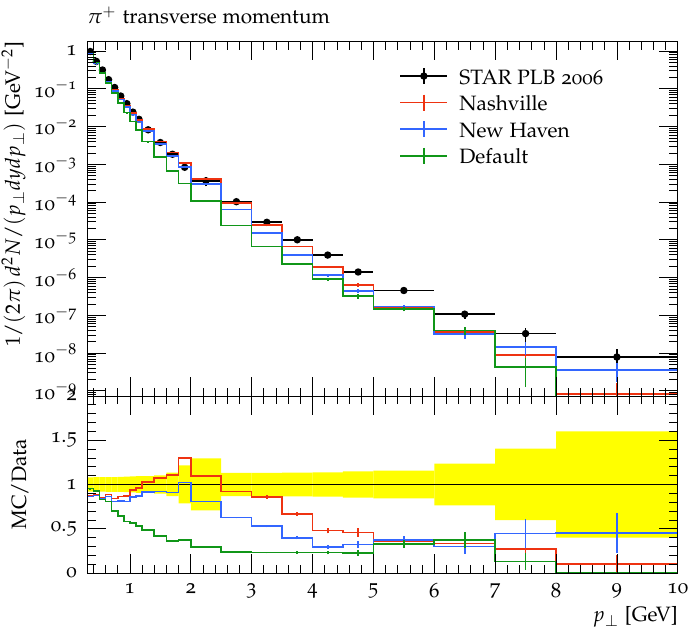}
    \end{minipage}
    \begin{minipage}{0.495\textwidth}
        \includegraphics[width=\linewidth]{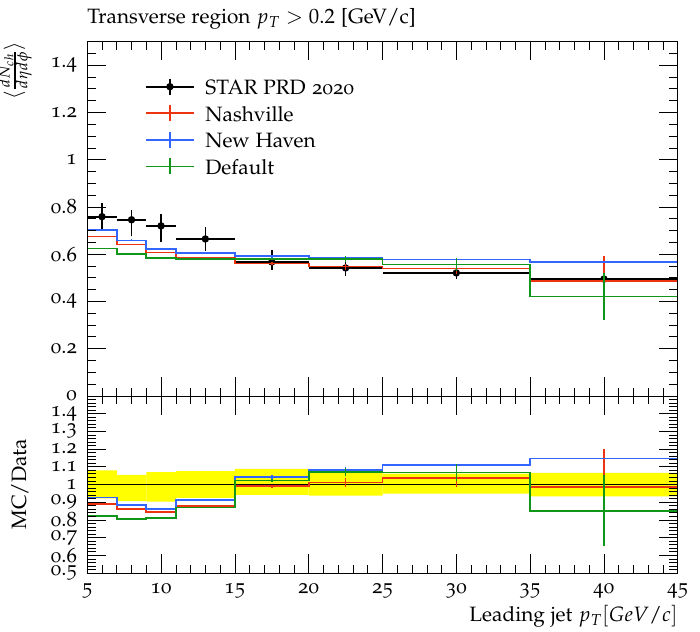}
    \end{minipage}
    
    \begin{minipage}{0.495\textwidth}
        \includegraphics[width=\linewidth]{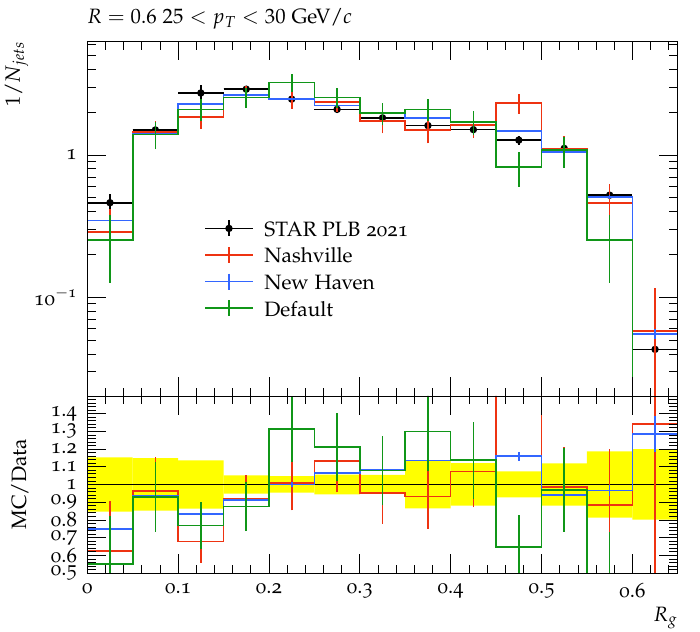}
    \end{minipage}
    \begin{minipage}{0.495\textwidth}
        \includegraphics[width=\linewidth]{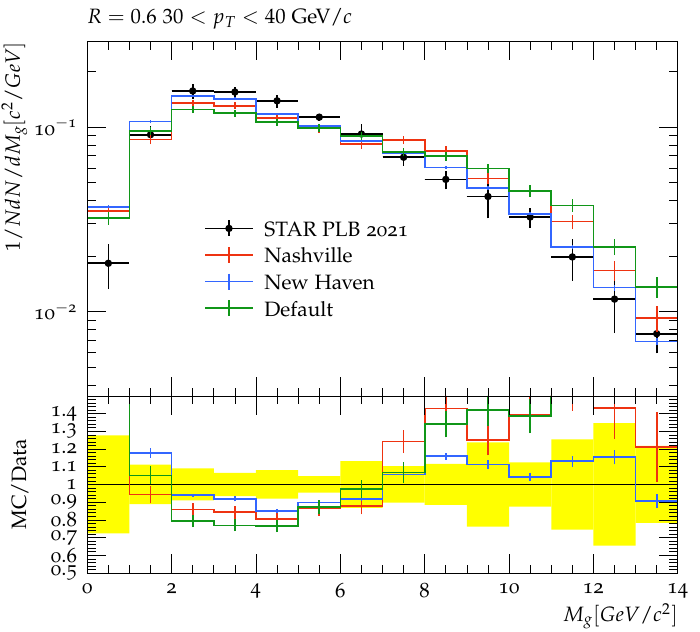}
    \end{minipage}
     \caption{Comparison of the default (green solid) and the new Nashville (red solid) and New Haven (blue solid) \textsc{Herwig}7 tunes with inclusive mid-rapidity $\pi^+$ invariant yields as a function of $p_\perp$ (top left), UE multiplicity as a function of leading jet $p_\perp$ (top right),  SoftDrop groomed jet radius $R_g$ (bottom left), and SoftDrop groomed jet mass $M_g$ (bottom right) from $pp$ collisions at $\sqrt{s}=200$ GeV as measured by the STAR experiment. The bottom panels in each plot show the ratio of the Monte Carlo predictions to measured data and the yellow-shaded region indicates the experimental uncertainties.}
    \label{fig:STAR_comps}
\end{figure*}

\begin{figure*}
    \centering
    \begin{minipage}{0.495\textwidth}
        \includegraphics[width=\linewidth]{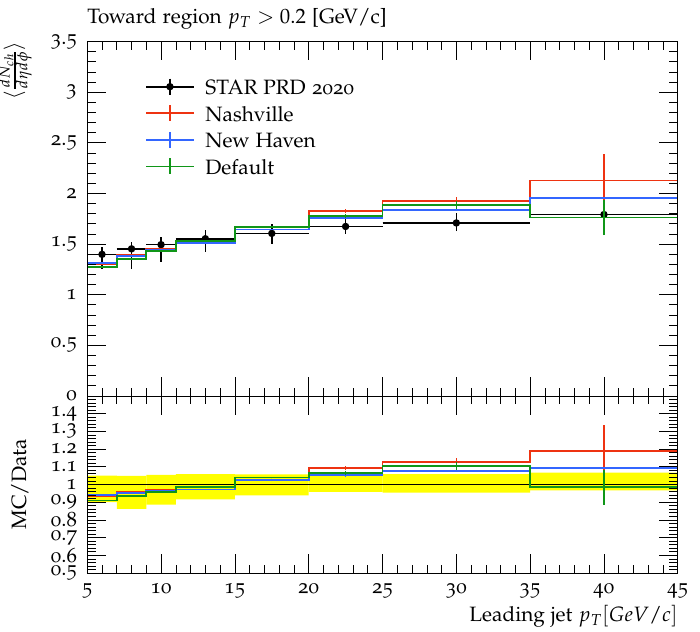}
    \end{minipage}
    \begin{minipage}{0.495\textwidth}
        \includegraphics[width=\linewidth]{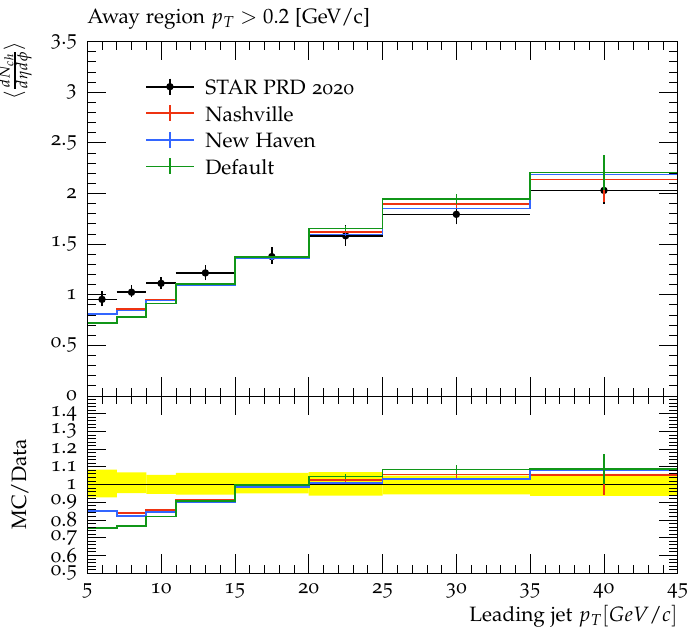}
    \end{minipage}
    
    \begin{minipage}{0.495\textwidth}
        \includegraphics[width=\linewidth]{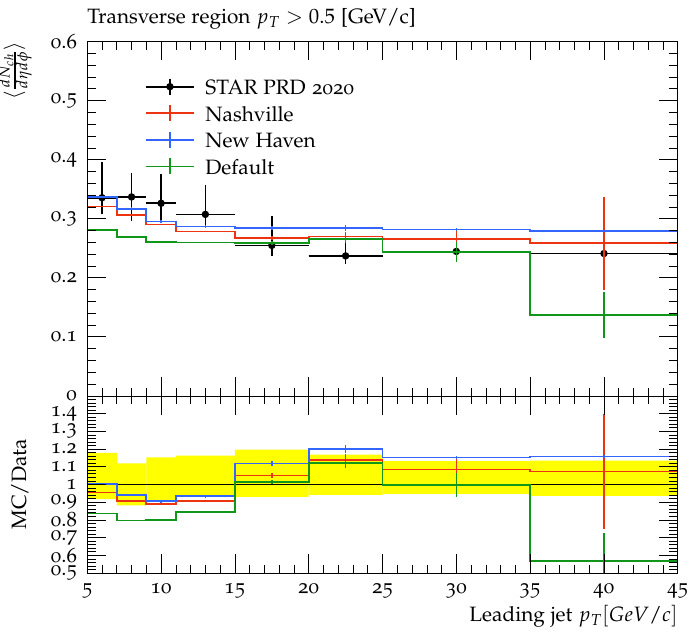}
    \end{minipage}
    \begin{minipage}{0.495\textwidth}
        \includegraphics[width=\linewidth]{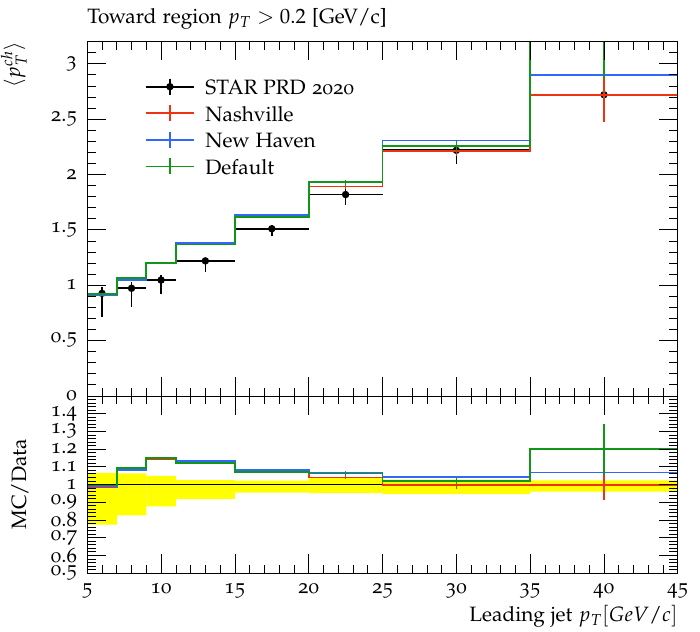}
    \end{minipage}
    \caption{UE observables as a function of leading jet $p_T$ from STAR measurements in proton-proton collisions at $\sqrt{s}=200$ GeV. The top left and right show the charge particle density in the towards and away regions, respectively. The bottom left and right figures show the charged particle density and average $p_T$ for the transverse and toward regions, respectively. The bottom panels in each plot show the ratio of the Monte Carlo predictions to measured data and the yellow-shaded region indicates the experimental uncertainties.}
    \label{fig:STAR_2019_UE}
\end{figure*}

\begin{figure*}
    \centering
    \begin{minipage}{0.495\textwidth}
        \includegraphics[width=\linewidth]{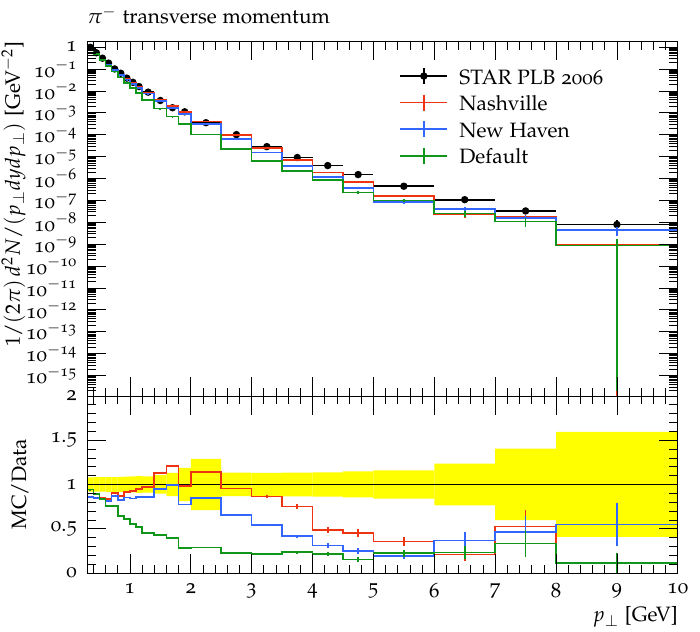}
    \end{minipage}
    \begin{minipage}{0.495\textwidth}
        \includegraphics[width=\linewidth]{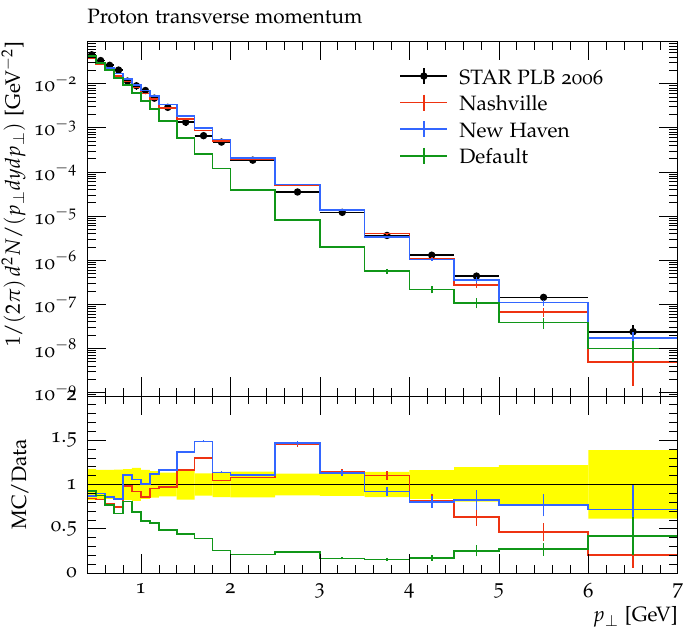}
    \end{minipage}
    \caption{Invariant yields of $\pi^-$ (left) and $p$ (right) as a function of $p_\perp$ from STAR measurements in proton-proton collisions at $\sqrt{s}=200$ GeV. The bottom panels in each plot show the ratio of the Monte Carlo predictions to measured data and the yellow-shaded region indicates the experimental uncertainties.}
        \label{fig:STAR_2006_PID}
\end{figure*}

\begin{figure*}
    \centering
    \begin{minipage}{0.495\textwidth}
        \includegraphics[width=\linewidth]{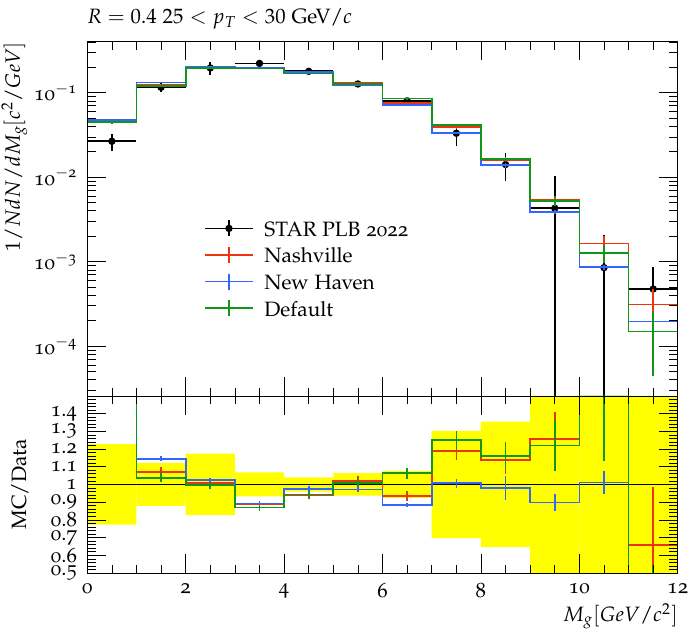}
    \end{minipage}
    \begin{minipage}{0.495\textwidth}
        \includegraphics[width=\linewidth]{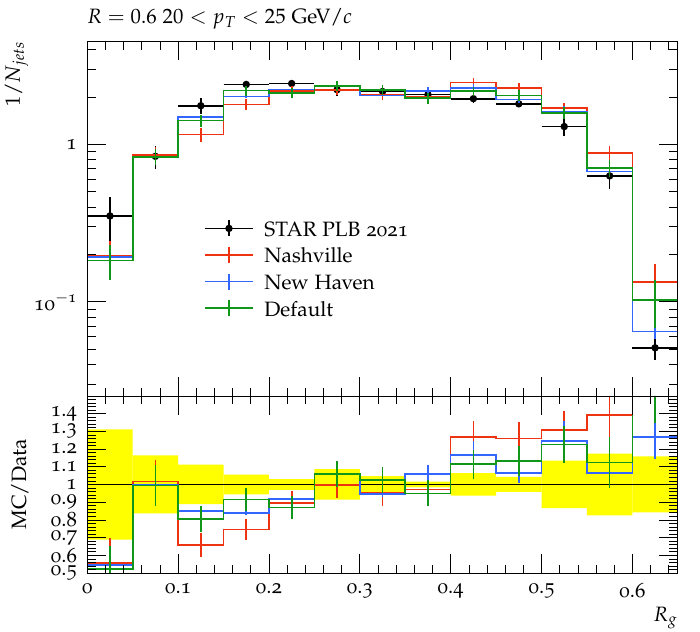}
    \end{minipage}
    \caption{Distributions of the SoftDrop groomed jet mass $M_g$ (left) and jet radius $R_g$ (right) from STAR measurements in proton-proton collisions at $\sqrt{s}=200$ GeV. The bottom panels in each plot show the ratio of the Monte Carlo predictions to measured data and the yellow-shaded region indicates the experimental uncertainties.}
    \label{fig:softDropgroomedjet}
\end{figure*}

\begin{figure*}
    \centering
    \begin{minipage}{0.495\textwidth}
        \includegraphics[width=\linewidth]{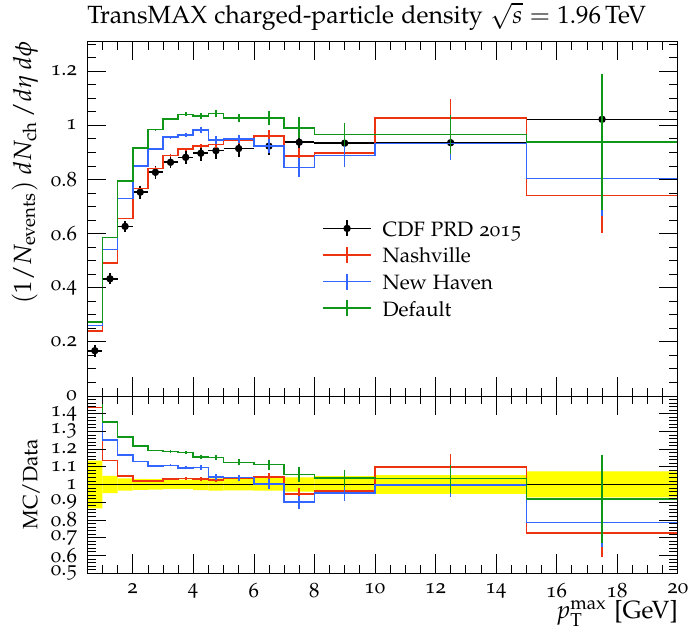}
    \end{minipage}
    \begin{minipage}{0.495\textwidth}
        \includegraphics[width=\linewidth]{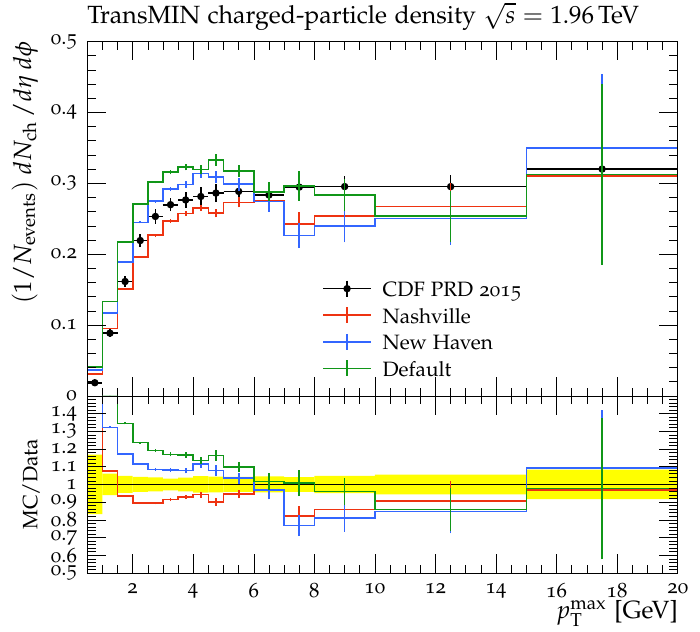}
    \end{minipage}
    \begin{minipage}{0.495\textwidth}
        \includegraphics[width=\linewidth]{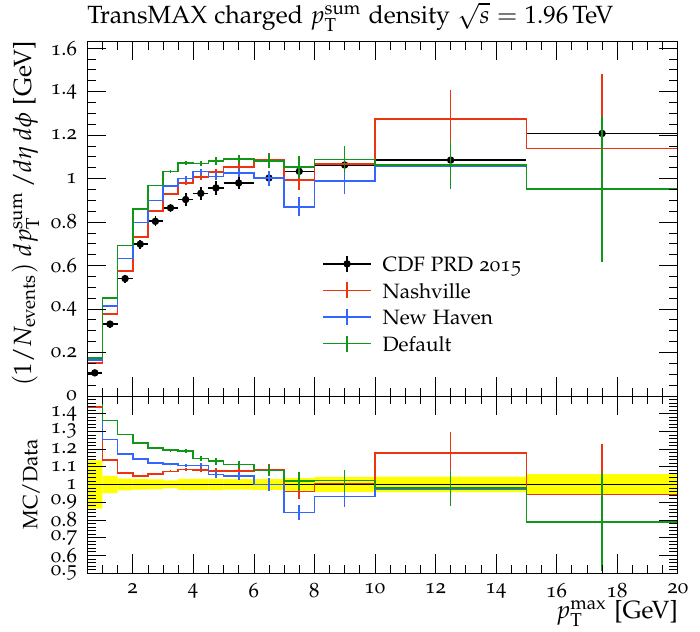}
    \end{minipage}
    \begin{minipage}{0.495\textwidth}
        \includegraphics[width=\linewidth]{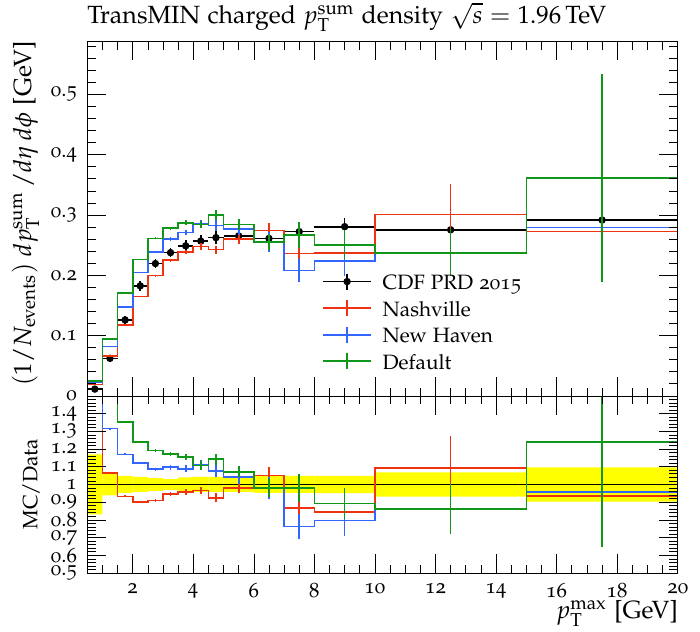}
    \end{minipage}
    \caption{UE observables as a function of leading hadron $p_T$ from CDF measurements in proton-antiproton collisions at $\sqrt{s}=1.96$ TeV. The top left and right show the charge particle multiplicity in the transMAX and transMIN regions (see text for definitions), respectively. The bottom left and right figures show the charged particle $p_T$ sum for the transMAX and transMIN regions, respectively. The bottom panels in each plot show the ratio of the Monte Carlo predictions to measured data and the yellow-shaded region indicates the experimental uncertainties.}
    \label{fig:CDF_1960}
\end{figure*}

\begin{figure*}
    \centering
    \begin{minipage}{0.495\textwidth}
    \includegraphics[width=\linewidth]{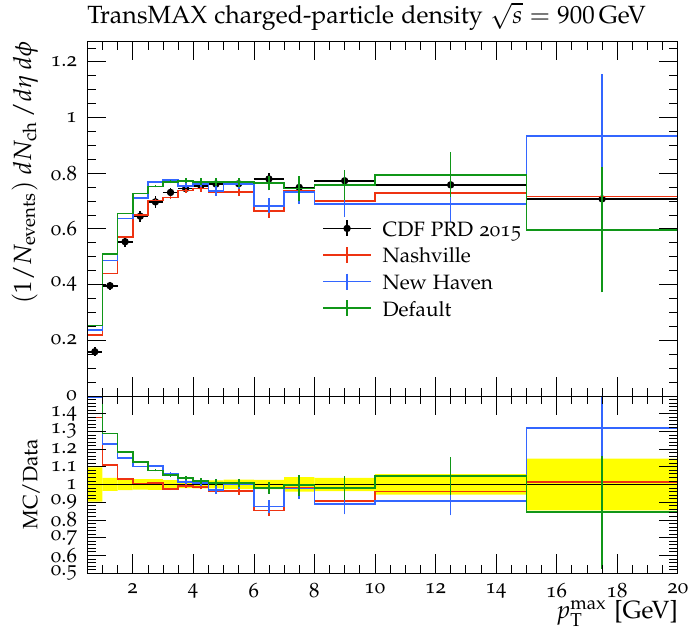}
    \end{minipage}
    \begin{minipage}{0.495\textwidth}
    \includegraphics[width=\linewidth]{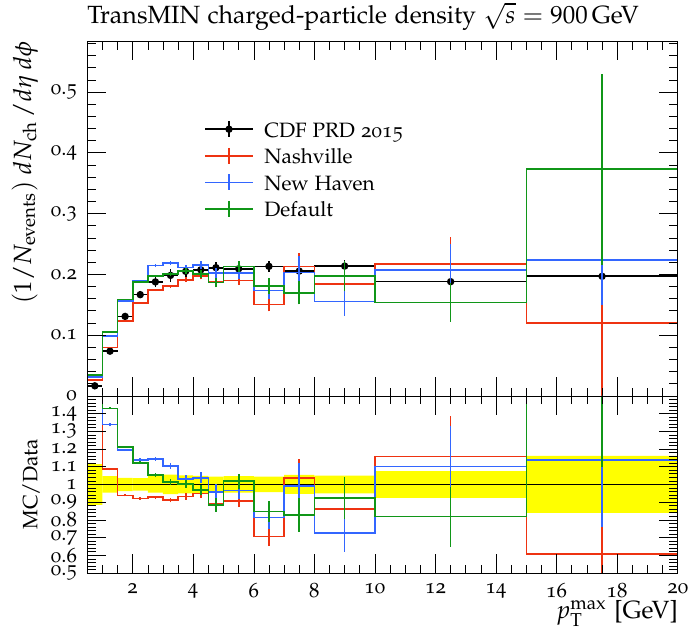}
    \end{minipage}
    
    \begin{minipage}{0.495\textwidth}
        \includegraphics[width=\linewidth]{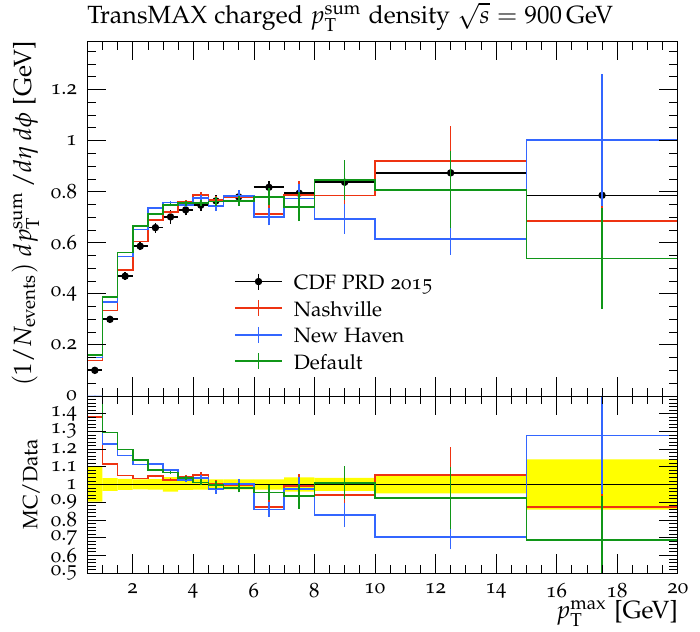}
    \end{minipage}
    \begin{minipage}{0.495\textwidth}
        \includegraphics[width=\linewidth]{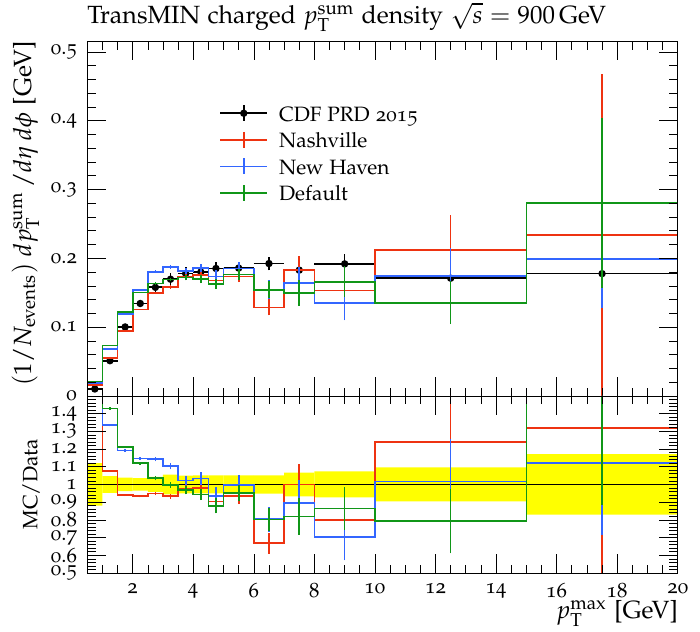}
    \end{minipage}
    \caption{UE observables as a function of leading hadron $p_T$ from CDF measurements in proton-antiproton collisions at $\sqrt{s}=900$ GeV. The top left and right show the charge particle multiplicity in the transMAX and transMIN regions (see text for definitions), respectively. The bottom left and right figures show the charged particle $p_T$ sum for the transMAX and transMIN regions, respectively. The bottom panels in each plot show the ratio of the Monte Carlo predictions to measured data and the yellow-shaded region indicates the experimental uncertainties.}
    \label{fig:CDF_900}
\end{figure*}

\begin{figure*}
    \centering
    \begin{minipage}{0.495\textwidth}
        \includegraphics[width=\linewidth]{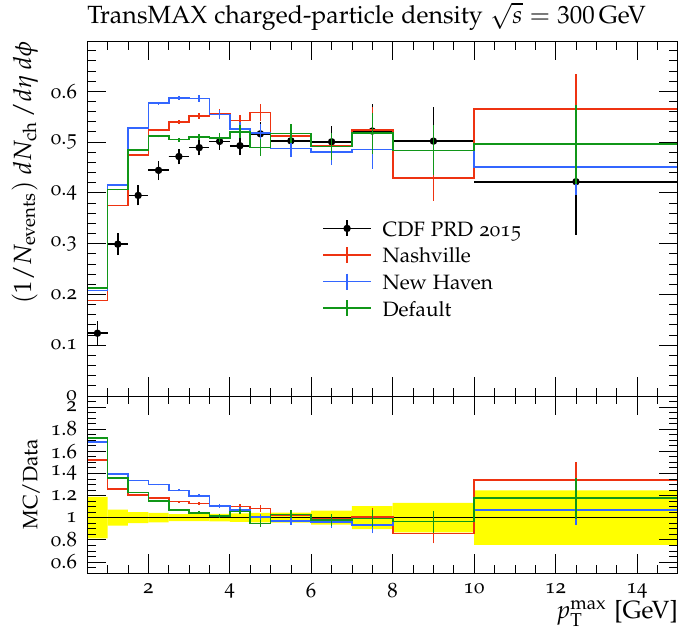}
    \end{minipage}
    \begin{minipage}{0.495\textwidth}
        \includegraphics[width=\linewidth]{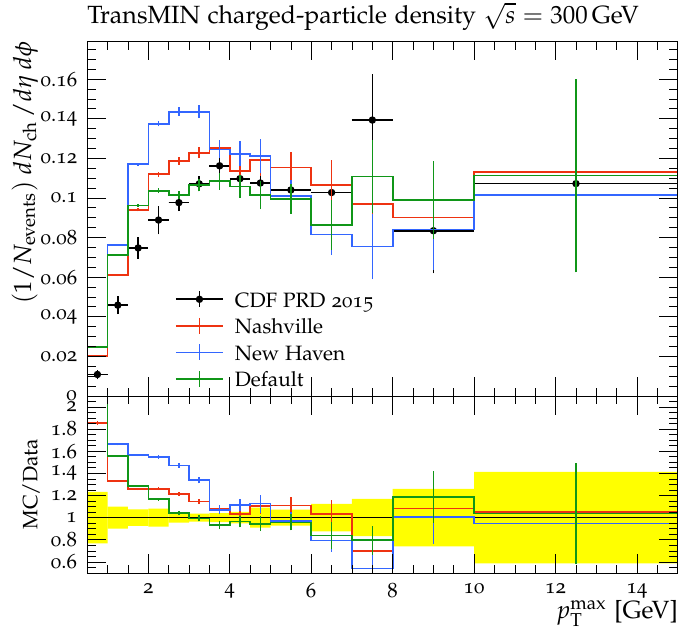}
    \end{minipage}
    
    \begin{minipage}{0.495\textwidth}
        \includegraphics[width=\linewidth]{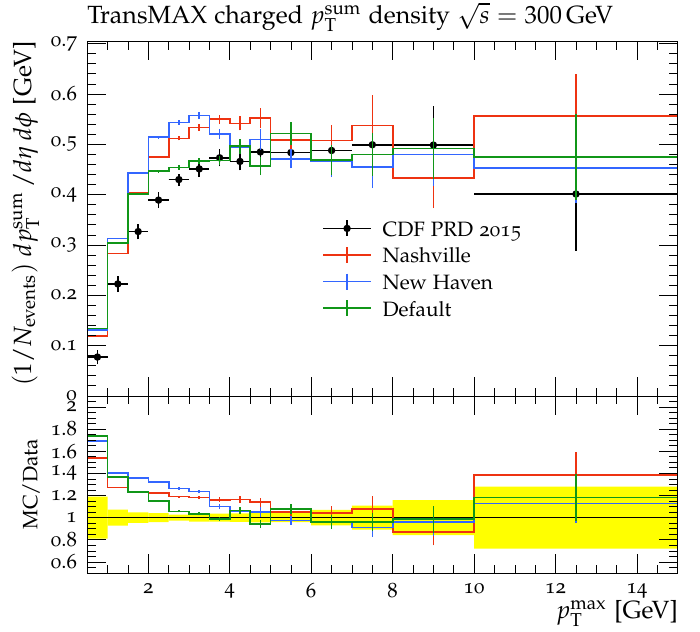}
    \end{minipage}
    \begin{minipage}{0.495\textwidth}
        \includegraphics[width=\linewidth]{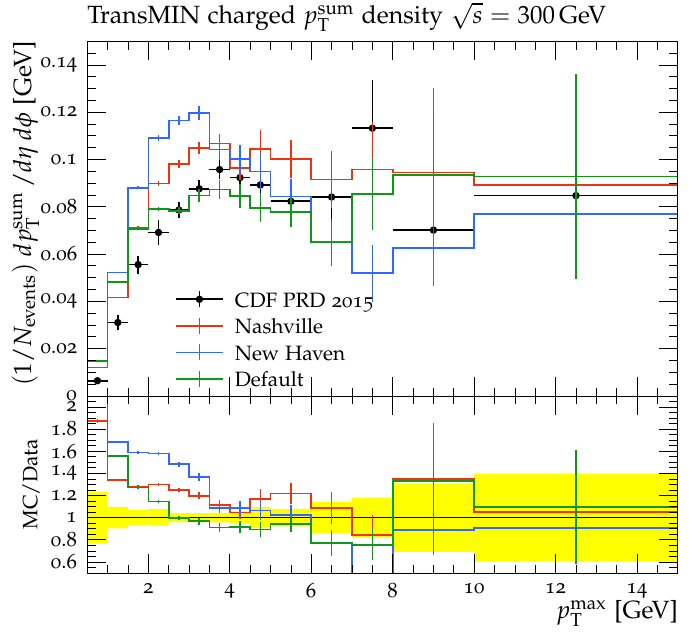}
    \end{minipage}
    \caption{UE observables as a function of leading hadron $p_T$ from CDF measurements in proton-antiproton collisions at $\sqrt{s}=300$ GeV. The top left and right show the charge particle multiplicity in the transMAX and transMIN regions (see text for definitions), respectively. The bottom left and right figures show the charged particle $p_T$ sum for the transMAX and transMIN regions, respectively. The bottom panels in each plot show the ratio of the Monte Carlo predictions to measured data and the yellow-shaded region indicates the experimental uncertainties.}
    \label{fig:CDF_300}
\end{figure*}

\begin{figure*}
    \centering
    \begin{minipage}{0.495\textwidth}
        \includegraphics[width=\linewidth]{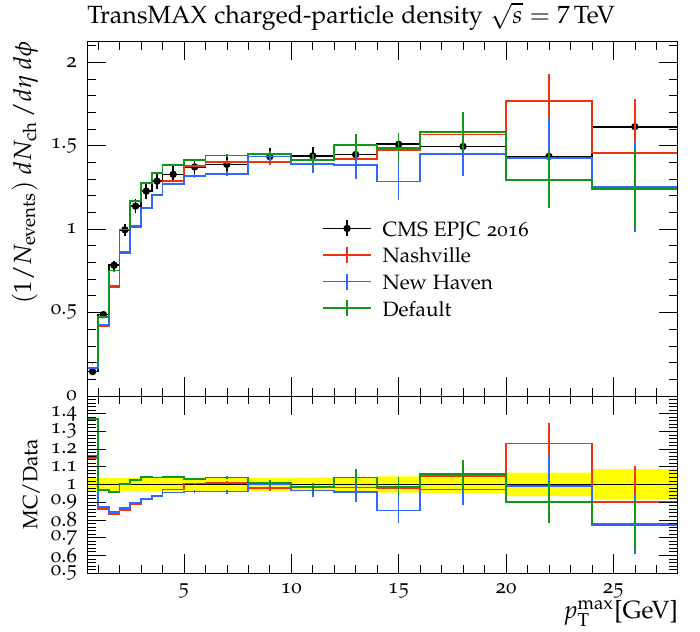}
    \end{minipage}
    \begin{minipage}{0.495\textwidth}
        \includegraphics[width=\linewidth]{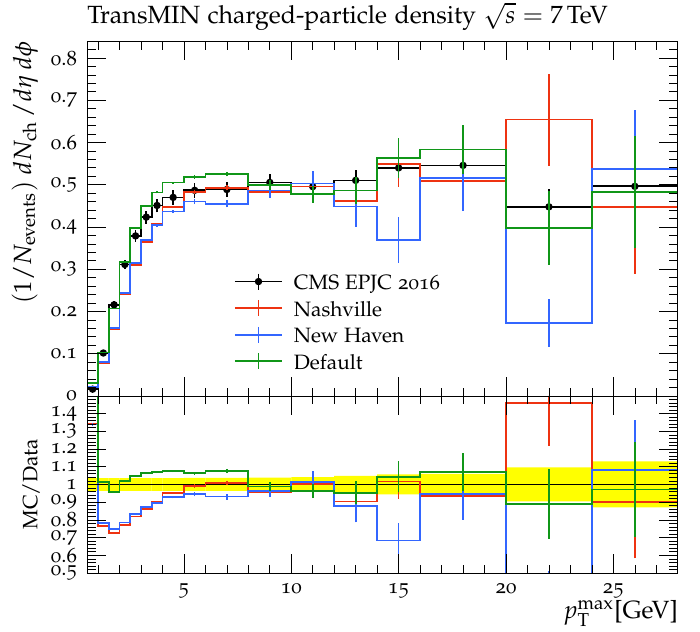}
    \end{minipage}
    
    \begin{minipage}{0.495\textwidth}
        \includegraphics[width=\linewidth]{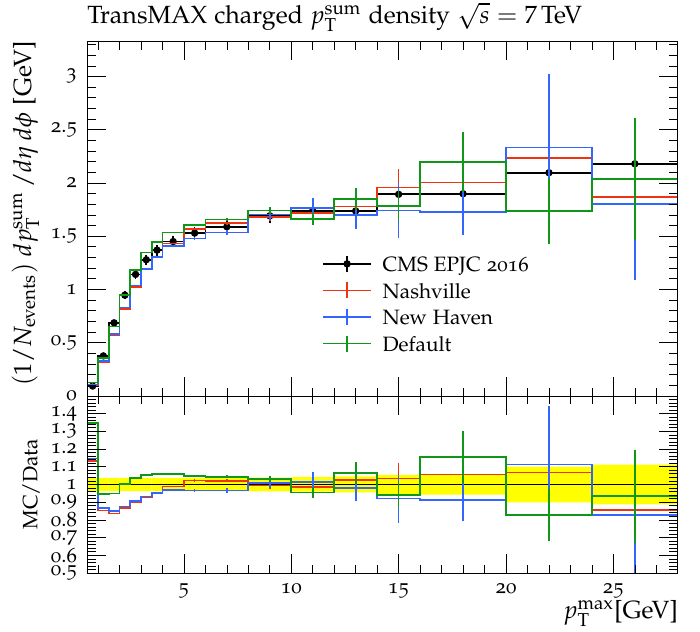}
    \end{minipage}
    \begin{minipage}{0.495\textwidth}
        \includegraphics[width=\linewidth]{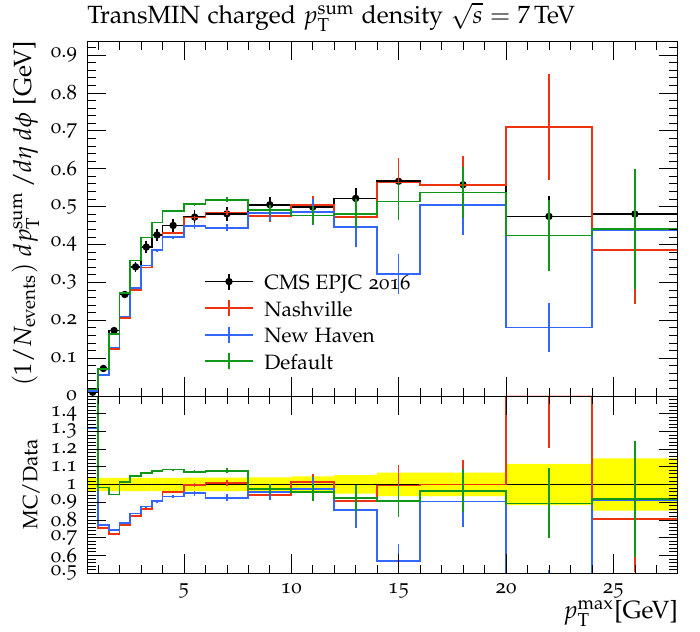}
    \end{minipage}
    \caption{UE observables as a function of leading hadron $p_T$ from CMS measurements in proton-proton collisions at $\sqrt{s}=7$ TeV. The top left and right show the charge particle multiplicity in the transMAX and transMIN regions (see text for definitions), respectively. The bottom left and right figures show the charged particle $p_T$ sum for the transMAX and transMIN regions, respectively. The bottom panels in each plot show the ratio of the Monte Carlo predictions to measured data, and the yellow-shaded region indicates the experimental uncertainties.}
    \label{fig:CMS_7000}
\end{figure*}

Figure \ref{fig:STAR_comps}, shows a comparison of the new Nashville and New Haven tunes with the default tune for a representative subset of the STAR data. The $\pi^+$ spectrum from $pp$ collisions at $\sqrt{s}=200$ (top left) GeV demonstrates varying levels of agreement with different tunes across the transverse momentum ($p_{\perp}$) range. At low $p_{\perp}$ (below $\sim 2$ GeV), the Nashville (red) and New Haven (blue) tunes align well with the data, though the Nashville tune slightly overestimates, while the default tune (green) underpredicts the spectrum. In the mid-$p_{\perp}$ range ($\sim$ 2 to $4$ GeV), Nashville tune offers the best agreement, closely following the data. However, New Haven and the default tune continue to underpredict the spectrum. At higher $p_{\perp}$ values (above 4 GeV), all three tunes increasingly deviate from the data, particularly beyond 5 GeV, where they all underestimate the measurements. Overall, the Nashville tune provides the best overall agreement across the full $p_{\perp}$ range, with the smallest deviations in the mid-$p_{\perp}$ region, while New Haven and the default tune show more significant discrepancies at both mid and high $p_{\perp}$.

The UE average charged particle density, as a function of the leading jet $p_{T}$ in the transverse region (top right of Figure \ref{fig:STAR_comps}), is generally well described by all three tunes across the full $p_{T}$ range, with only minor discrepancies in the low-$p_{T}$ ($\sim$ 5 to 10 GeV) and high-$p_{T}$ ($\gtrsim 30$ GeV) regions. In the low-$p_T$ region, in particular, while the discrepancy is small for all three tunes, the default tune underpredicts the data slightly more than New Haven and Nashville. In the mid-$p_{T}$ range ($\sim$ 10 to 30 GeV), Nashville and New Haven tunes continue to track closely with the data, showing excellent agreement. The Nashville tune provides the most consistent agreement across the entire leading jet $p_{T}$ range, with New Haven offering similarly good performance. The default tune slightly underestimates the data in some bins but remains within reasonable bounds. Overall, all three tunes provide satisfactory modeling of the underlying event.

All three tunes are reasonably close to the SoftDrop groomed jet radius $R_g$ data (bottom left of Figure \ref{fig:STAR_comps}) below $R_g = 0.2$, staying well within experimental uncertainties. As $R_g$ increases (between $\sim$ 0.2 and 0.4), the Nashville tune remains most closely aligned with measurement; the New Haven and default tunes start to slightly overestimate the data, though the New Haven tune does so more modestly. At higher $R_g$ values (above 0.4), all three tunes show slight deviations from the data, with the default tune deviating the most and overestimating the groomed jet radius. Overall, in this plot and throughout all observed plots of $R_g$, the New Haven and Nashville tunes provide better agreement across the full $R_g$ range than the default tune; in this particular plot, the Nashville tune performs best overall.

With respect to the SoftDrop groomed jet mass distribution for jets in the $p_T$ range of 30 to 40 GeV (bottom right of Figure \ref{fig:STAR_comps}), all three tunes exhibit a noticeable overprediction in the first bin (below 1 GeV) compared to the data. As the groomed jet mass increases from 1 to 4 GeV, New Haven and Nashville tunes align more closely with the data, while the default tune underpredicts slightly but stays within the uncertainty limits. In the mid- ($\sim$ 4 to 10 GeV) and high-$M_g$ ($\gtrsim$ 10 GeV) regions, New Haven continues to provide the best match to the data, with the other two tunes showing a slight overprediction and the default tune deviating more significantly. The Nashville tune's deviations lessen at the highest $M_g$ bins ($\sim 12$ to 14 GeV). Overall, New Haven offers the most consistent description of the data, while Nashville performs comparably well, especially at lower masses, and the default tune shows more discrepancies, particularly in the mid-to-high jet mass range.

Figure \ref{fig:STAR_2019_UE} compares the underlying event (UE) observable distributions in the Toward, Away, and Transverse regions. In the top left plot, the distribution of leading jet $p_T$ in the Toward region ($p_T > 0.2$ GeV) is shown. All three tunes follow the data reasonably well across the entire $p_T$ range. The top right plot shows the leading jet $p_T$ distribution in the Away region ($p_T > 0.2$ GeV). Here, all three tunes generally follow the data closely; there is slight underprediction by all tunes at low leading jet $p_T$, with the default tune deviating the most from the data points.

In the bottom left plot, the distribution of the leading jet $p_T$ in the Transverse region ($p_T > 0.5$ GeV) is shown. The Nashville and New Haven tune provides the best agreement across the $p_T$ range, although the Nashville tune follows the data better. At low and high leading jet $p_T$, the default tune deviates from the data, more drastically in the latter region than the former.

Finally, the bottom right plot shows the average transverse momentum density, $\langle p_T^{\text{avg}} \rangle$, in the Toward region ($p_T > 0.2$ GeV). All three tunes follow the data closely, with slight deviations at low leading jet $p_T$ and a larger deviation in the default tune at the highest $p_T$ bin.

Figure \ref{fig:STAR_2006_PID} shows the $\pi^{-}$ meson and proton transverse momentum spectra at $\sqrt{s} = 200$ GeV. In the $\pi^{-}$ distribution (left), the Nashville tune provides the best overall agreement with the data across the entire $p_{\perp}$ range. The New Haven tune also follows the data reasonably well but deviates slightly more than Nashville, especially where $2 \lesssim p_{\perp} \lesssim 6$. At the lowest transverse momentum ($p_{\perp} \lesssim 1$ GeV), all three tunes agree with data. Otherwise, the default tune consistently underpredicts, showing noticeable discrepancies from the data throughout the $p_{\perp}$ spectrum. At higher $p_{\perp}$ values (above 4 GeV), all tunes begin to diverge significantly from the data. In the proton transverse momentum distribution (right), similar trends are observed, though the New Haven tune remains closely aligned with data even up to the highest $p_{\perp}$ bins.

Figure \ref{fig:softDropgroomedjet} compares the SoftDrop groomed jet mass $M_g$ (left) and groomed radius $R_g$ (right) distributions. In the jet mass distribution (left), all three tunes show reasonable agreement with the data across the full $M_g$ range, with the New Haven tune in closest agreement. In the groomed radius distribution (right), all tunes exhibit similar behavior, tracking the data well. The Nashville Tune tends to slightly underestimate the data at lower $R_g$ and overpredict at higher $R_g$, while New Haven Tune provides the better overall agreement across the full range. 

Figure \ref{fig:CDF_1960} illustrates underlying event (UE) observables from CDF measurements in $p\overline{p}$ collisions at $\sqrt{s}=1.96$ TeV. A similar trend is observed in both the TransMAX and TransMIN regions, with respect to both the charged-particle density and charged $p_T^{\text{sum}}$ density: all three tunes tend to lie relatively close to the data and experimental uncertainty, except at low $p_T^{\text{max}}$. In this region ($p_T^{\text{max}} \lesssim 6$ GeV), New Haven and the default tune overpredict the data somewhat, while Nashville shows the closest agreement, only very slightly overpredicting (underpredicting) in the TransMAX (TransMIN) regions. Additionally, there is slight underprediction from both Nashville and New Haven in the region $6 \lesssim p_T^{\text{max}} \lesssim 10$ in all observables, perhaps moreso in the TransMIN region. Overall, however, we observe that New Haven and Nashville provide a better description of the UE data recorded by CMS at $\sqrt{s} = 1.96$ TeV than the default tune, with the best performance by the Nashville tune. Similar trends can be seen in Figure \ref{fig:CDF_900} and \ref{fig:CDF_300}, which shows UE observables from CDF measurements in $p\overline{p}$ collisions at $\sqrt{s}=900$ and $\sqrt{s}=300$ GeV. At $\sqrt{s} = 900$ GeV (Figure \ref{fig:CDF_900}), Nashville maintains the best performance out of the three tunes, with New Haven and the default tune performing similarly, particularly at low $p_T^{\text{max}}$. As we decrease the center-of-mass energy further to $\sqrt{s} = 300$ GeV (Figure \ref{fig:CDF_300}), Nashville performs best in the very lowest $ p_T^{\text{max}} $ bins (up to $\sim 2$ GeV), while the default tune performs better in the intermediate region. For $ p_T^{\text{max}} \gtrsim 6 $ GeV, all three tunes converge and perform similarly well compared to the data.

Finally, Figure \ref{fig:CMS_7000} shows UE observables from CMS measurements in $pp$ collisions at $\sqrt{s}=7$ TeV. In the lowest $ p_T^{\text{max}} \lesssim 3 $ GeV bins, the default tune demonstrates excellent agreement with the data, outperforming the other tunes. However, this improved agreement in the low-$ p_T $ region comes at the expense of overshooting the data at intermediate $ p_T $ values (approximately $ 3 \, \text{GeV} \lesssim p_T^{\text{max}} \lesssim 5 \, \text{GeV} $, or up to ~7 GeV for transMIN). For $ p_T^{\text{max}} \gtrsim 7 \, \text{GeV} $, all tunes converge and show good agreement with the data.

\section{Discussion}\label{sec:discussion}
This study presents new underlying event (UE) tunes for the \textsc{Herwig}7.3 Monte Carlo event generator, specifically developed to improve the description of UE observables at RHIC energies while maintaining strong performance at higher energies, such as those probed by CDF and CMS. The default \textsc{Herwig}7 tune, designed primarily for high-energy data, struggles with predicting particle production at lower energies like those at RHIC. This study introduces two new tunes, Nashville and New Haven, that address this limitation by better capturing the dynamics at these lower energies.

The tuning process involved fitting the parameters of the \textsc{Herwig}7 model using the Professor toolkit, utilizing a $\chi^2$-based fit to experimental data from STAR, PHENIX, CDF, and CMS. The Nashville tune was developed primarily based on Tevatron and LHC data, but also incorporates RHIC data, offering a comprehensive tune that improves accuracy across a wide range of energies with a trade-off on particle spectra yields at RHIC. The New Haven tune, meanwhile, was tailored more specifically to RHIC conditions, ensuring enhanced performance at $\sqrt{s} = 200$ GeV.

Results demonstrate that at the nominal RHIC energy of $\sqrt{s} = 200$ GeV, both new tunes surpass the default \textsc{Herwig} tune, which tends to either overpredict or underpredict UE observables. At this center-of-mass energy, the Nashville tune appears to provide the greatest improvement in describing charged particle spectra, while the New Haven tune offers slightly better descriptions of jet substructure observables. Both tunes provide slightly better predictions of charged particle and charged $p_T$ densities at $\sqrt{s} = 200$ GeV when compared to the default tune.

At higher center-of-mass energies, the Nashville tune consistently provides the closest alignment with experimental data, especially in the TransMAX and TransMIN regions of charged particle density and $ p_T^{\text{sum}} $ density, across both low and high $ p_T^{\text{max}} $ ranges. The New Haven tune also performs well for these observables (albeit with slight deviations in some regions) given the fact these data were not used during the fitting process for the New Haven tune. It is, however, worth noting that the New Haven tune's ability to describe the data slightly worsens as one increases energy from 300 to 900 to 1960 GeV. This is not unexpected, as the New Haven tune was constructed only using data taken at $\sqrt{s} = 200$ GeV. At the highest center-of-mass energy seen here ($\sqrt{s} = 7$ TeV), all three tunes offer comparable performance, demonstrating the new tunes’ versatility. A future study involving the compatibility between the parameter sets and an updated model parameterization is underway with the aim of extracting a global minimum in the fitting procedure. While this study only considered data at mid-rapidity, it is important that future tunes include measurements at forward/backward-rapidity from forward detectors of experiments such as STAR, LHCb, and the upcoming Electron-Ion Collider. One expects the inclusion of forward data in tunes to necessitate a rapidity-dependent parameterization of MC models for particle production, as we move towards the universal description of scale variation between the hard and soft physics in event generation.


\section*{Acknowledgments}

The authors would like to thank Simon Platzer (Uni-Graz) and Stefan Gieseke (KITP) for helpful discussions on the various parameters and implementation within the \textsc{Herwig} UE model. The authors also thank John Lajoie (ORNL) for testing the new tune independently and for helpful feedback on the draft. RKE and USQ would like to thank the Vanderbilt ACCRE computing facility. RKE would like to acknowledge funding by the U.S. Department of Energy, Office of Science, Office of Nuclear Physics under grant number DE-SC0024660.
HLC, IAM, and LJM thank the Yale Center for Research Computing. HLC and IAM acknowledge funding by the Office of Nuclear Physics of the U.S. Department of Energy under award number DE-SC004168 (HLC and IAM) and BNL/DOE-424803 (IAM).

\nocite{*}

\bibliography{apssamp}

\end{document}